\documentclass{article}

\usepackage{PRIMEarxiv}

\usepackage[utf8]{inputenc} 
\usepackage[T1]{fontenc}    
\usepackage{hyperref}       
\usepackage{url}            
\usepackage{booktabs}       
\usepackage{amsfonts}       
\usepackage{nicefrac}       
\usepackage{microtype}      
\usepackage{lipsum}
\usepackage{fancyhdr}       
\usepackage{graphicx}       
\graphicspath{{media/}}     

\newcommand{\var}{\texttt}
\usepackage[ruled,vlined]{algorithm2e}
\usepackage{amsmath}

\title{Proactive Autoscaling for Edge Computing Systems with Kubernetes
\thanks{Accepted by 2021 IEEE/ACM 14th International
Conference on Utility and Cloud Computing (UCC ’21) Companion} 
}

\author{
  Li Ju, Prashant Singh, Salman Toor \\
  Department of Information Technology \\
  Uppsala University \\
  Uppsala, Sweden\\
  \texttt{\{li.ju, prashant.singh, salman.toor\}@it.uu.se} \\
}

\begin{document}
\maketitle

\begin{abstract}
With the emergence of the Internet of Things and 5G technologies, the edge computing paradigm is playing increasingly important roles with better availability, latency-control and performance. However, existing autoscaling tools for edge computing applications do not utilize heterogeneous resources of edge systems efficiently, leaving scope for performance improvement. In this work, we propose a Proactive Pod Autoscaler (PPA) for edge computing applications on Kubernetes. The proposed PPA is able to forecast workloads in advance with multiple user-defined/customized metrics and to scale edge computing applications up and down correspondingly. The PPA is optimized and evaluated on an example CPU-intensive edge computing application further. It can be concluded that the proposed PPA outperforms the default pod autoscaler of Kubernetes on both efficiency of resource utilization and application performance. The article also highlights future possible improvements on the proposed PPA. 
\end{abstract}


\section{Introduction}
Cloud computing applications have gained increasing popularity in various wide-ranging domains over the last few decades\cite{velte2010cloud, varghese2018next}. Cloud computing vendors provide their users on-demand availability of a vast amount of resources, including computing units, storage, network devices, and even services and applications. Cloud computing has proven to be successful in both commercial and technical aspects, and the approach powers a lot of popular applications (e.g. Netflix, DropBox, Spotify)\cite{shimba2010cloud, roumani2019empirical}. However, cloud computing platforms suffer from latency of applications, which is mainly limited by the geographic distances and bandwidths of the network between clients and data centres\cite{barker2010empirical}. Thus, cloud computing does not fully meet the needs of latency-sensitive applications (e.g. video game streaming, real-time data analysis). 

\begin{figure}[h]
    \centering
    \includegraphics[width=0.5\linewidth]{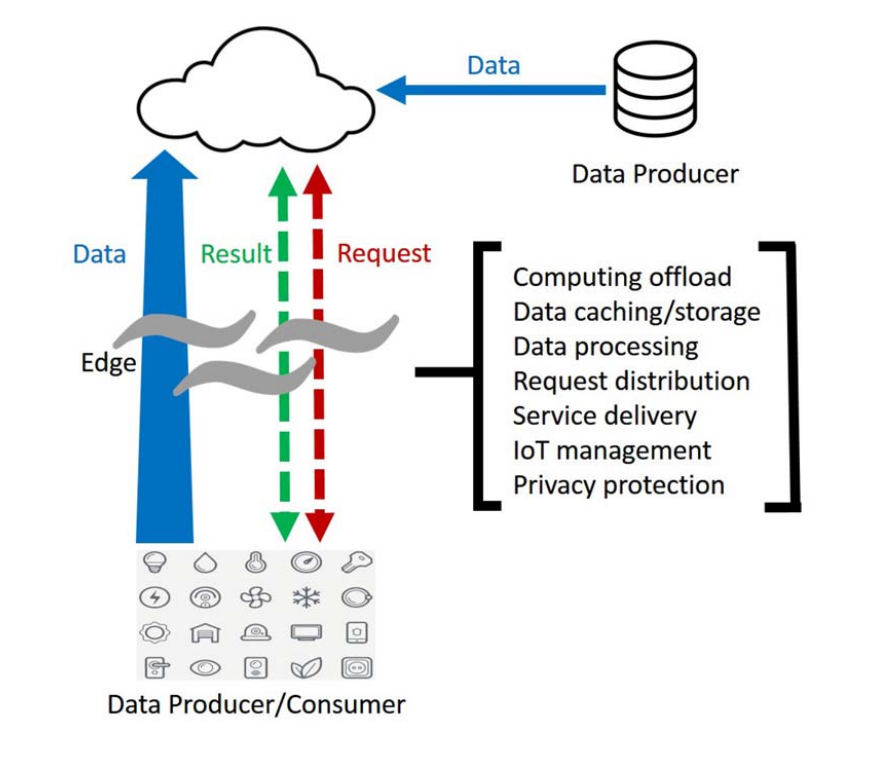}
    \caption{Edge computing paradigm\cite{shi2016promise}}
    \label{fig:edgecomputing}
\end{figure}

Edge computing is a promising approach to solve the latency issue of traditional cloud computing\cite{shi2016promise}, by moving latency-sensitive computation and data storage to locations closer to clients towards the edge of the network (shown as Figure \ref{fig:edgecomputing}). Featuring lower latency, bandwidth occupation and overheads, edge computing is playing important roles in realizing smart cities, smart grids, smart transportation, etc\cite{yu2017survey}. 

In real world scenarios, autoscaling of applications for both traditional cloud computing and edge computing is essential\cite{al2013impact, xu2017online}. With variation in workloads, autoscaling tools adjust the amount of resources dynamically and automatically, and keep a stable average workload for each computation/storage unit. Autoscaling provides better fault tolerance, high availability, efficient energy consumption and cost management for applications. 

For cloud computing, Horizontal Pod Autoscaler (HPA) is widely used as a native service provided by Kubernetes, the de facto cloud framework on most cloud platforms\cite{lorido2014review}. Given CPU utilization as the metric for workloads, HPA scales cloud applications in a reactive way with the algorithm presented in Equation \ref{eq:hpaalg}. 

\begin{equation}\label{eq:hpaalg}
    \mathrm{NumOfReplicas} = \mathrm{ceil} \left( \frac{\mathrm{CurrentMetricValue}}
    {\mathrm{PredefinedMetricValue}} \right)
\end{equation}

For edge computing, however, autoscaling is a more complicated and complex problem\cite{wang2017enorm}. Though HPA is simple and effective in many cloud computing cases, it is not fully capable to autoscale edge computing applications for three reasons: 
\begin{itemize}
    \item HPA is not specifically designed for edge computing environments, being unaware of constraints and capacities of heterogeneous edge resources; 
    \item HPA only considers CPU utilization to estimate workloads. However, for edge computing applications, other information about the system (e.g. job queues, I/O, request rates) is essential when scaling applications as well; 
    \item HPA is rule-based and unprogrammable, and service providers can hardly customize HPA to meet their specific needs for edge computing applications
\end{itemize}


A promising approach is to develop a proactive autoscaler for edge computing applications which supports multiple metrics and customizable algorithms. Developing such an autoscaler is challenging. Theoretically, due to the heterogeneity and limitations of edge resources, predicting workloads and scaling pods on edge systems form a mixed problem of time series analysis and multi-objective optimization with constraints. Practically, collecting multiple metrics from different sources and organizing custom algorithms depending on various libraries in a uniform way is an involved multi-framework engineering problem. 

In this work, we propose a multi-metrics supported and customizable proactive pod autoscaler. The autoscaler is able to collect multiple metrics, predict workloads and scale target applications in advance. Moreover, it allows users to customize their own scaling policies and predicting models to better fits their applications. The main contributions of this work include: 
\begin{enumerate}
    \item Introducing a proactive workflow for autoscaling edge computing applications.
    \item Implementing the multi-metrics supported and customizable autoscaler on Kubernetes.
    \item Conducting experiments to optimize and evaluate the proposed autoscaler.
\end{enumerate}

The remainder of this article is organized as follows. Related literature is reviewed in Section 2. In Section 3, the design of the edge system environment we use in this work are illustrated. In Section 4, the architecture and implementation of the proposed autoscalers is explained in detail. Section 5 and 6 present experimental designs and results for the proposed autoscaler. Section 7 concludes the work, and highlight further possible improvements. 

\section{Related Work}
Considering HPA as a baseline, a number of research works have been pursued to explore alternatives to HPA for cloud/edge applications hosted on containers. In this section, related worsk about both reactive and proactive autoscaling techniques will be reviewed. Also, the necessity of our proposed Proactive Pod Autoscaler and its unique features are explained. 
\subsection{Reactive Autoscalers}
In 2017, Yahya et al. presented a framework named ELASTICDOCKER to vertically scale running containers reactively \cite{al2017autonomic}. With predefined thresholds, ELASTICDOCKER is able to monitor CPU and memory usage and assign additional CPU and memory resources to target containers if they exceed predefined thresholds. However, vertical scaling is limited by physical resources and entails greater risks of outages and hardware failures, which brings unreliability in real-world use. 

Many reactive horizontal autoscalers are developed as well, to improve performance and flexibility of HPA. In 2019, Fan et al. reported a system architecture for container autoscaling and conducted experiments to evaluate their autoscalers\cite{zhang2019quantifying}. Compared with HPA, its algorithm is more complex and efficient. In 2020, an article from Salman and Marko explored other key factors that autoscalers on Kubernetes should take into consideration beside the CPU utilization\cite{taherizadeh2020key} to improve the Quality of Services. It is confirmed that multiple metrics are necessary to better autoscale cloud systems, in addition to CPU utilization. 

\subsection{Proactive Autoscalers}
A major disadvantage of reactive autoscalers is the delay of control. Though containers will be scaled once workloads have changed, initialization or termination of containers takes time. To solve the issue, a promising idea is to forecast workloads and make scaling decisions in advance. This is named as proactive/predictive autoscaling and there are a few related articles reported. 

In 2016, Yang et al. presented a predictive autoscaling algorithm named CRUPA for container resource utilization based on time series analysis\cite{meng2016crupa}. Auto Regressive Integrated Moving Average models are used for predicting CPU usage of containers, and based on predicted values, the same provision algorithm as HPA (\ref{eq:hpaalg}) is used. Another example is from Tian et al., reporting another predictive autoscaling framework for containers in 2017 \cite{ye2017auto}. A hybrid autoscaling strategy combining prediction of resources utilization and Service Level Agreement (SLA) is proposed to provide better Quality of Services. The single metric they selected for autoscaling is CPU utilization, and the predictive model is Auto Regressive Moving Average model. 

Machine learning models are gaining a lot of attention in analysis of time series data\cite{siami2018comparison} and there are some proactive autoscalers based on machine learning models as well. In 2020, Mahmoud, Imtiaz and Mohammad proposed a machine learning-based proactive autoscaler, which utilizes LSTM and multiple metrics to predict workloads\cite{imdoukh2019machine}. The autoscaler is validated on a real-world access dataset Worldcup98. 

Though there are a few works about proactive autoscaling for cloud systems, proactive autoscalers for edge computing systems are rare. To the best of our knowledge, there is only one article reported proposing proactive autoscaling for edge systems\cite{abdullah2020predictive}. A pretrained neural network model is prepared and used as the predictive model in the running system to forecast CPU utilization. Predicted CPU metrics are used to estimate number of replicas the system requires. 

\begin{table*}
\begin{center}
\begin{tabular}{c|c|c|c|c}
\hline
    \textbf{Authors} & \textbf{Strategy} & \textbf{Metrics} & \textbf{Systems} & \textbf{Models}\\
\hline
    Fan et al.\cite{zhang2019quantifying}    & Reactive  & CPU                      & Cloud & -\\
    Salman and Marko\cite{taherizadeh2020key} & Reactive  & CPU, loop interval, etc. & Cloud & -\\
    Yang et al.\cite{meng2016crupa}     & Proactive & CPU                      & Cloud & ARIMA\\
    Tian et al.\cite{ye2017auto}      & Proactive & CPU \& SLA               & Cloud & ARMA\\
    Mahmoud et al.\cite{siami2018comparison} & Proactive & CPU, RAM \& Requests & Cloud & LSTM\\
    Muhammad et al.\cite{imdoukh2019machine}     & Proactive & CPU               & Edge & EN, DTR\\
    \textbf{Proposed}              & Proactive & CPU, RAM, I/O \& Custom Metrics & Edge & Custom Models\\
\hline
\end{tabular}
\caption{Comparison of related work and proposed method}
\label{tab:litrev}
\end{center}
\end{table*}

Table \ref{tab:litrev} compares related work and the method we are proposing. Our proposed Proactive Pod Autoscaler (PPA) features following points: 
\begin{itemize}
    \item it is one of few studies which focuses on autoscaling for edge computing applications considering limitations and constraints of heterogeneous resources. 
    \item multiple metrics (CPU, RAM, I/O utilization) and custom metrics are supported to autoscale applications, providing alternative metrics for workload estimation and forecasting. Though CPU utilization is capable to predict workloads in many cases, some applications require multiple, or application-specific metrics to make scaling decisions.
    \item the predictive model for the PPA is highly customizable. Unlike other proactive pod autoscalers with fixed predicting models, the proposed PPA supports custom models and multiple model frameworks (e.g. \var {TensorFlow}, \var {statsmodels}, etc.). Users can inject their own models into the PPA to best suit their needs and to obtain optimal performance for their applications. 
\end{itemize}


\section{System Design}
The setup of the considered cloud and edge environments is now presented in detail.
\subsection{Edge Computing Environment}\label{sec:sectionlabel}
\begin{figure}[h]
    \centering
    \includegraphics[width=0.8\linewidth]{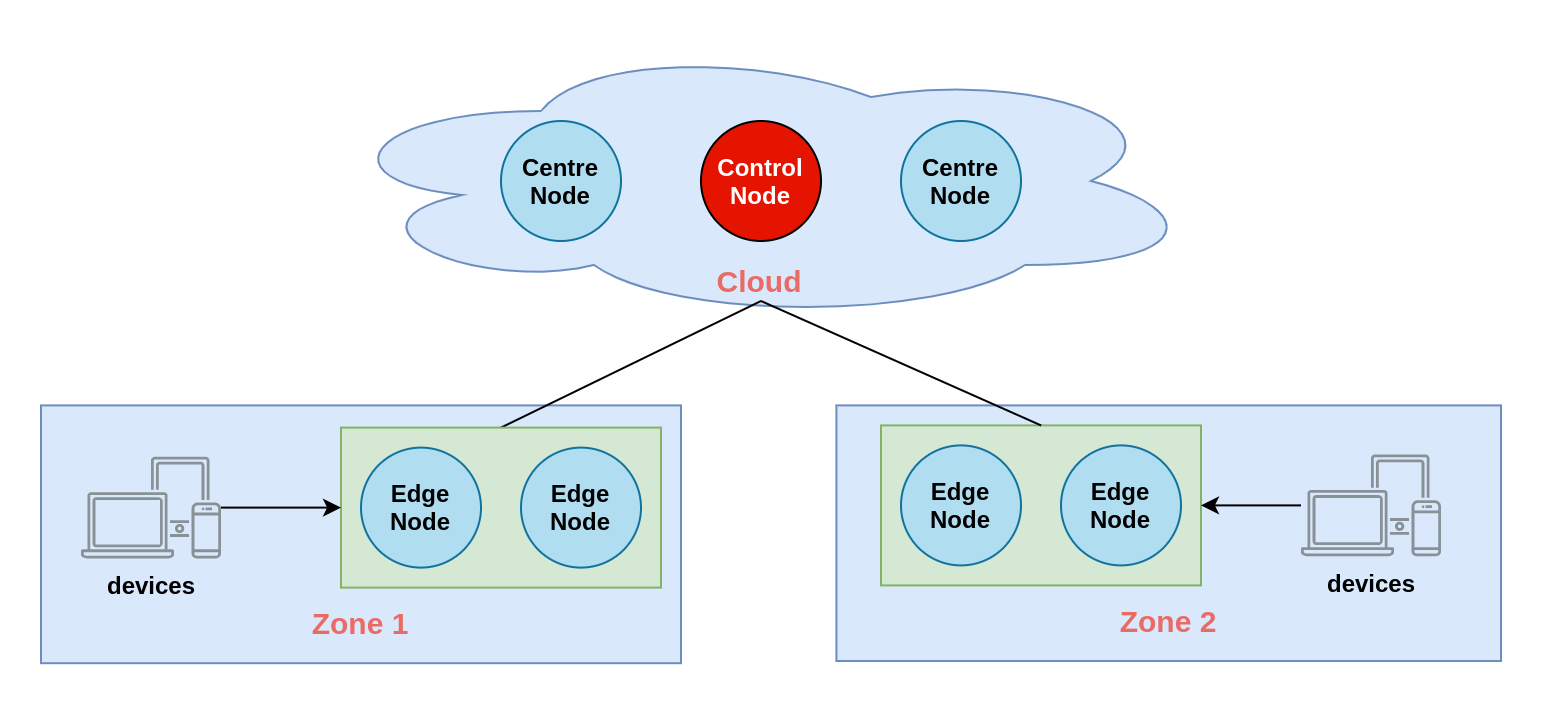}
    \caption{Topology of the edge computing system}
    \label{fig:clusterTop}
\end{figure}

Figure \ref{fig:clusterTop} shows the topology we have orchestrated for the cloud/edge computing environment. Nodes in each of the environments are connected by Kubernetes. The described system depicts a real-world model of a typical cloud/edge computing environment. Table \ref{tab:cluster} introduces assigned resources of nodes in cloud, and at the edge. The allocated resources at the edge environment are limited and heterogeneous,  whereas the cloud environment hosts stronger computing capacities and larger memory size resources. The information regarding software and frameworks used for the deployment and execution are listed in Table \ref{tab:sysinfo}. 

\begin{table}[h]
\begin{center}
\begin{tabular}{c|c|c|c|c}
\hline
    \textbf{Role} & \textbf{Tier} & \textbf{CPU/millicores} & \textbf{RAM/GB} & \textbf{Number}\\
\hline
    Control & Cloud & 4000 & 4 & 1\\
    Worker  & Cloud & 3000 & 3 & 2\\
    Worker  & Edge  & 2000 & 2 & 2/zone\\
\hline
\end{tabular}
\caption{Physical resources of nodes in cloud and at edge}
\label{tab:cluster}
\end{center}
\end{table}

\begin{table}[h] \begin{center}
\begin{tabular}{c|c}
\hline
    \textbf{Name} & \textbf{Version}\\
\hline
    Ubuntu & 20.04.2\\
    Kubectl & v1.20.2\\
    Docker  & 20.10.2\\
    Prometheus stack & 16.1.2\\
    Custom Pod Autoscaler & v1.1.0\\
    Celery & 5.1.0\\
\hline
\end{tabular}
\caption{Information of software used in the work}
\label{tab:sysinfo}
\end{center}
\end{table}

\subsection{Framework Components \& Metrics}

\begin{figure}[h]
    \centering
    \includegraphics[width=0.5\linewidth]{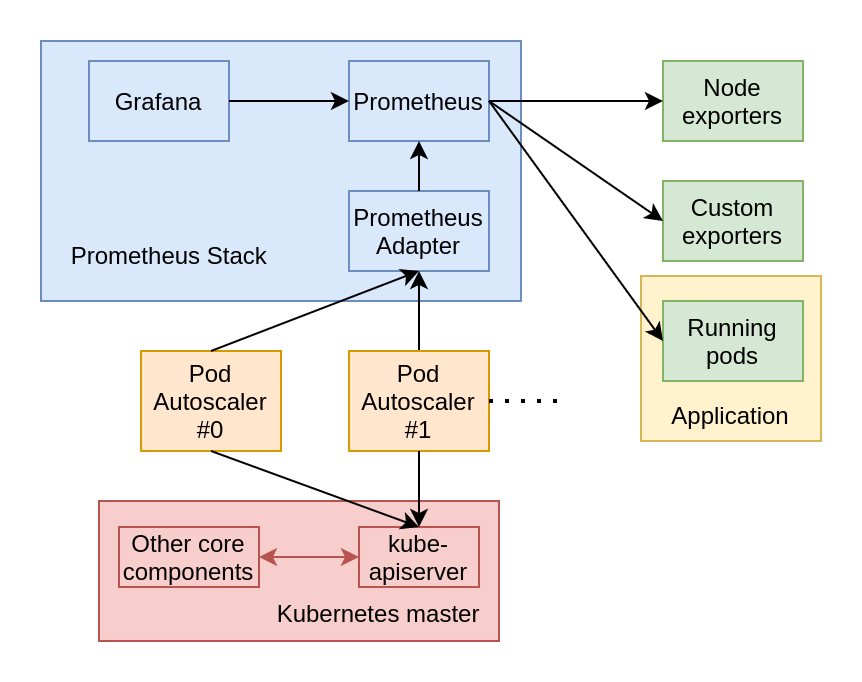}
    \caption{Logic components and metrics flow of the cluster}
    \label{fig:monitoring}
\end{figure}

The monitoring system plays an essential role towards successfully running the proposed cloud/edge computing framework. It is responsible for regularly collecting running metrics from other components, exposing collected metrics to other internal parts, and preparing a comprehensive view of the entire system. For monitoring, we rely on the already available production-grade monitoring services. Our approach is to use modular services that can easily be replaced by other components in future. Within the scope of the proposed framework, the monitoring system is based on the Prometheus stack\cite{prometheus_stack}, which is one of the most popular frameworks for the cloud and edge computing environments. Each logic component of the environment and flow of metrics are introduced as following. 

\subsubsection{Exporters}
Exporters of different types are responsible for providing different metrics for Prometheus. In the considered edge computing environment, node exporters are deployed on each node for node-level metrics, and a custom exporter is deployed for application-specific metrics. 

\subsubsection{Prometheus stack}
Prometheus stack consists of three parts - Prometheus, Grafana and Prometheus Adapter. Prometheus is responsible for collecting metrics from exporters in an active way (i.e. pulling metrics instead of being pushed). Collected metrics are visualized using Grafana, and are exposed by the Prometheus Adapter within the cluster in a standard API of Kubernetes, so that other pods are able to fetch them for internal uses. 

\subsubsection{Autoscalers}
From the Prometheus Adapter, autoscalers fetch all types of required metrics, evaluate the number of desired replicas, and make requests for scaling decisions to the Kubernetes master control panel. It is Kubernetes' responsibility to handle scaling requests and schedule new pods on nodes. 

To better autoscale worker pods across all zones, autoscalers are deployed in the cloud zone instead of edges closer to their target pods. There are 2 reasons for this: 
\begin{itemize}
    \item Both the Prometheus stack and the control panel of Kubernetes are deployed in the cloud zone. Autoscalers in the cloud zone enjoy less latency when fetching metrics and when making scaling requests;  
    \item The proactive autoscalers may need to update neural network models, which entails heavy computation resources. Placing autoscalers in the cloud ensures enough computation capacity. 
\end{itemize}

\section{Architecture \& Implementation}
In this section, the architecture and the implementation details of the proposed Proactive Pod Autoscaler are introduced. Firstly the structure and workflow of the PPA are explained, providing a general understanding of how the PPA works. Thereafter, detailed algorithms of each component of the PPA are motivated and illustrated. The implementation of the proposed PPA is finally explained, including global settings of the PPA that can be configured, and interfaces of models based on different libraries/frameworks. 
\subsection{Structure \& Workflow}
\begin{figure}[h]
    \centering
    \includegraphics[width=0.8\linewidth]{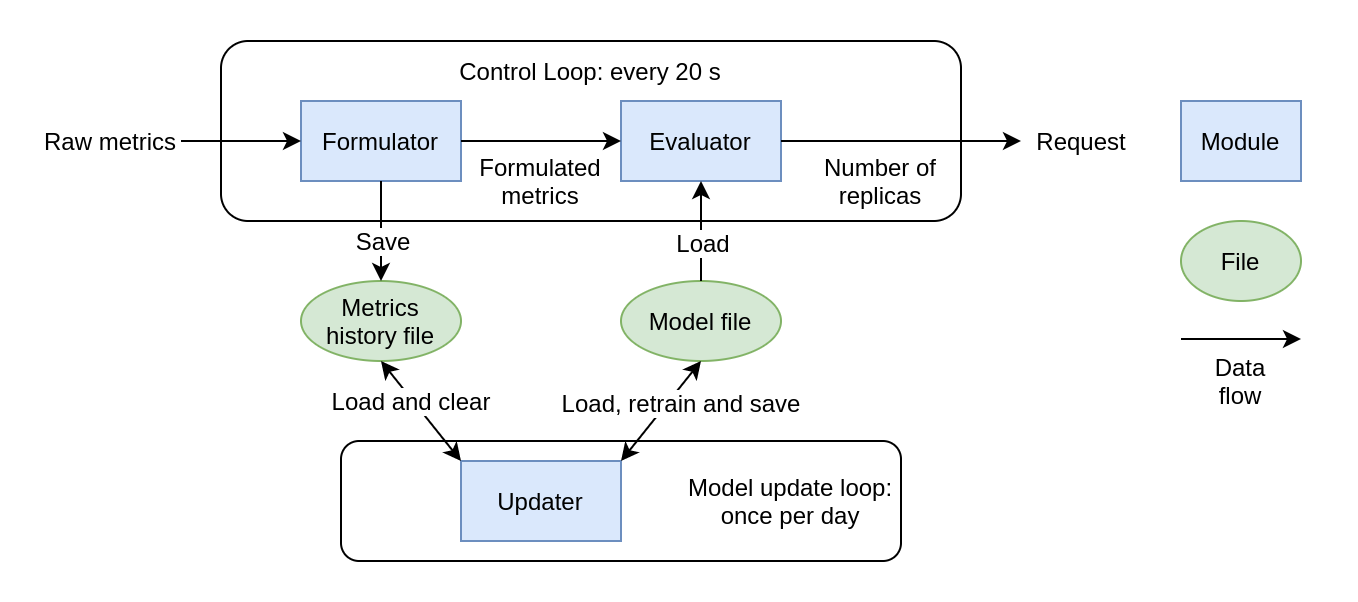}
    \caption{Architecture of the proposed autoscaler}
    \label{fig:cpaarc}
\end{figure}

Figure \ref{fig:cpaarc} shows the structure of the proposed proactive pod autoscaler. The PPA consists of three components (\emph{Formulator}, \emph{Evaluator} and \emph{Updater}), works in two loops (\emph{control loop} and \emph{model update loop}), and maintains two files (\emph{metrics history file} and \emph{model file}). The initialization of the PPA requires a pretrained seed model to be injected in the framework as an initial \emph{model file}. 

\subsubsection{Control Loop}
In each \emph{control loop}, the PPA is responsible for scaling the target pods using the collected metrics. Raw metrics will firstly be fetched from the Prometheus Adapter and passed to the \emph{Formulator}. After extraction from raw data, required metrics are stored in the \emph{metrics history file} and passed along to the \emph{Evaluator} by the \emph{Formulator}. The \emph{Evaluator} loads the model from the \emph{model file}, and predicts the number of desired replicas with the formulated metrics. Finally the request for a given scaling decision is made to Kubernetes control panel. 

\subsubsection{Model Update Loop}
In each \emph{model update loop}, the \emph{Updater} loads data from the \emph{metrics history file} as the training set and update the model. After the model has been updated, the \emph{Updater} will remove the \emph{metrics history file} and re-save the model to the \emph{model file}. With the \emph{Updater}, models for prediction are continually updated and improved for new workload patterns.

\subsection{Algorithms}

\subsubsection{Evaluator}
Algorithm \ref{alg:eval} described the workings of the \emph{Evaluator}. For each PPA, a \emph{key metric} must be set as an estimator for workloads, and injected predictive models will be used to predict the key metric. To arrive at the number of desired replicas from the predicted key metric, a \emph{Static Policy} should be defined. 

In this work, the threshold based algorithm of HPA described in Section \ref{eq:hpaalg} is used as the default \emph{Static Policy}. Still, \emph{Static Policies} are customizable and users may inject their own policies in the PPA. 

The algorithm ensures following features of PPA: 
\begin{enumerate}
    \item \textbf{Proactive:} Predictive models are used to forecast workloads based on trends identified using current metrics, which ensures the PPA to autoscale applications in a proactive way. 
    \item \textbf{Limitation-aware:} The PPA is aware of resource limitations of target nodes, and it will not overscale the application resulting in exceeding physical limitations. 
    \item \textbf{Robust} The PPA will make requests based on current metrics even if it fails to predict future metrics, which may happen because the \emph{model file} is being updated, or if the file is corrupted by accident. 
    \item \textbf{Model-diagnostic:} As long as the injected model meets model protocols, PPA is able to use it for evaluation and updating. This ensures users much flexibility to use any models which fits their applications. 
    \item \textbf{Confidence-considered:} If injected models are Bayesian, which are able to produce confidence/uncertainty estimates of each prediction, the PPA is able to utilize the confidence of prediction. The PPA works in a proactive way only if the prediction of models is confident enough over the preset confidence threshold, otherwise the PPA makes decisions on based on current metrics. 
\end{enumerate}

\begin{algorithm}
\SetAlgoLined
\KwResult{Number of replicas to be requested}
Get \var {current\_metrics}\;
Calculate \var {max\_replicas} limited by system resources\;
\var {model} $\leftarrow$ \var {Load} (\var {model\_file})\;
\eIf{\var {model.isValid()}}
    {\var {key\_metric} $\leftarrow $ \var {Predict} (\var {model}, \var {current\_metrics})\;
    \If{\var {model.isBayesian()} \textbf{and} \var {prediction\_confidence} $<$ \var {confidence\_threshold}}
        {\var {key\_metric} $\leftarrow$ \var {current\_key\_metric}}}
    {\var {key\_metric} $\leftarrow$ \var {current\_key\_metric}}
\var {num\_replicas} $\leftarrow$ \var {Static\_Policies} (\var {key\_metric})\;
\If{\var {num\_replicas} $>$ \var {max\_replicas}}
    {\var {num\_replicas} $\leftarrow$ \var {max\_replicas}}
 \caption{Algorithm for the \emph{Evaluator}}
 \label{alg:eval}
\end{algorithm}

\subsubsection{Protocol of Models}
Though the PPA is flexible, allowing users to inject customized time series models, protocols must be followed by the models as defined below:  
\begin{enumerate}
    \item \textbf{Size of time window:} In this work, we define the size of time window for all models as 1 unit, which indicates that models are predicting workload of next 1 \emph{control loop} using metrics of last 1 loop. This is decided by the initialization time cost for new pods, which generally takes less than one time interval of \emph{control loops}. 
    \item \textbf{Input \& Output metrics:} Input metrics of models should be listed as following \var {[CPU, RAM, Network Input, Network Output, Custom Metric]}. The model should predict all input variables but only one metric is set as the key metric. 
\end{enumerate}

\subsubsection{Model Update Policy of the Updater}\label{sec:updatePolicy}
Though the workflow of the \emph{Updater} is fixed, we proposed 3 different policies to update the seed model (referred as \emph{model update policies}): 
\begin{enumerate}
    \item \textbf{Do not retrain the model:} The injected seed model is updated and is used throughout execution. This approach is proposed due to the large cost of updating a model, especially for deep neural network models. If patterns of workloads are stable, and the seed model is able to produce comparable results with those produced by updated models, it is not necessary to update model with large amount of resources periodically;
    \item \textbf{Retrain a model from the scratch:} In each \emph{model update loop}, the \emph{Updater} loads the training set, drops the old model and trains a new model from scratch with the same architecture as the seed model. In cases where workloads of different days vary substantially, models based on old data do not necessarily fit workload patterns of coming days and a new model is a good choice; 
    \item \textbf{Update the model:} The old model are retrained with data of last \emph{model update loop} for several extra epochs with the training set. In many cases, workloads of patterns do change but not by much, and it is a good choice to use the old model as a starting point for the update process. 
\end{enumerate}
For different applications and different workload patterns, different approaches may be optimal. Here in this work, experiments are designed and conducted to compare the 3 approaches to find the optimal performance on the example applications. 

\subsection{Implementation}
The proposed PPA is implemented based on a third-party extension for Kubernetes named Custom Pod Autoscaler (CPA)\cite{cpa2020jamie}. CPA runs as a container, and the implementation of the PPA involves development of a standard Docker Image. The CPA framework provides a standard protocol that allows users to develop their own autoscalers as long as it outputs the desired number of replicas in a JSON file. In this work, the PPA is implemented in Python with necessary libraries. 

All user customizable arguments for the PPA are passed using environment variables when creating the container and available arguments are listed in Table \ref{tab:customargs}. 

\begin{table}[h]
\centering
\begin{tabular}{c|c|c}
\hline
    \textbf{Arg} & \textbf{Type} & \textbf{Descriptions}\\
\hline
    ModelLink & str & Link of the custom model \\
    ScalerLink  & str & Link of the scaler\\
    ModelType  & str & Type of the model \\
    KeyMetric & str & The key metric to be predicted\\
    ControlInterval & int & Period of \emph{control loops} (/s)\\
    UpdateInterval & int & Period of \emph{model update loops} (/h)\\
    CustomExporter & str & Link of the custom exporter\\
    Threashold & float & Threashold on the key metric\\
\hline
\end{tabular}
\caption{Configurable arguments for the proposed PPA}
\label{tab:customargs}
\end{table}

The proposed PPA is designed to be model-agnostic. To use models of different types and frameworks in a uniform way, a standard helper interface is designed as an abstract class. With link and type of the custom model specified, the proposed autoscaler fetches the model automatically and creates a corresponding helper object to deal with models in the initialization stage. The helper object ensures different models can be loaded, updated, saved and predicted with in a consistent way. Currently, 2 helper classes have been implemented for libraries \var {Keras}\cite{chollet2015keras} and \var {statsmodels}\cite{seabold2010statsmodels}. If custom models are created by other libraries (e.g. \var {PyTorch}), users can implement their own helper classes following protocols of the helper interface. 

\section{Experiments}
The goal of the experiments is to optimize the proposed Proactive Pod Autoscaler for an example application, and to validate its performance in comparison to HPA.
In this section, firstly the example application and generation of workloads are introduced. Then details of experiments for the optimization and evaluation for the PPA are explained. 

\subsection{Example Application}\label{sec:app} 
\begin{figure}[h]
    \centering
    \includegraphics[width=0.5\linewidth]{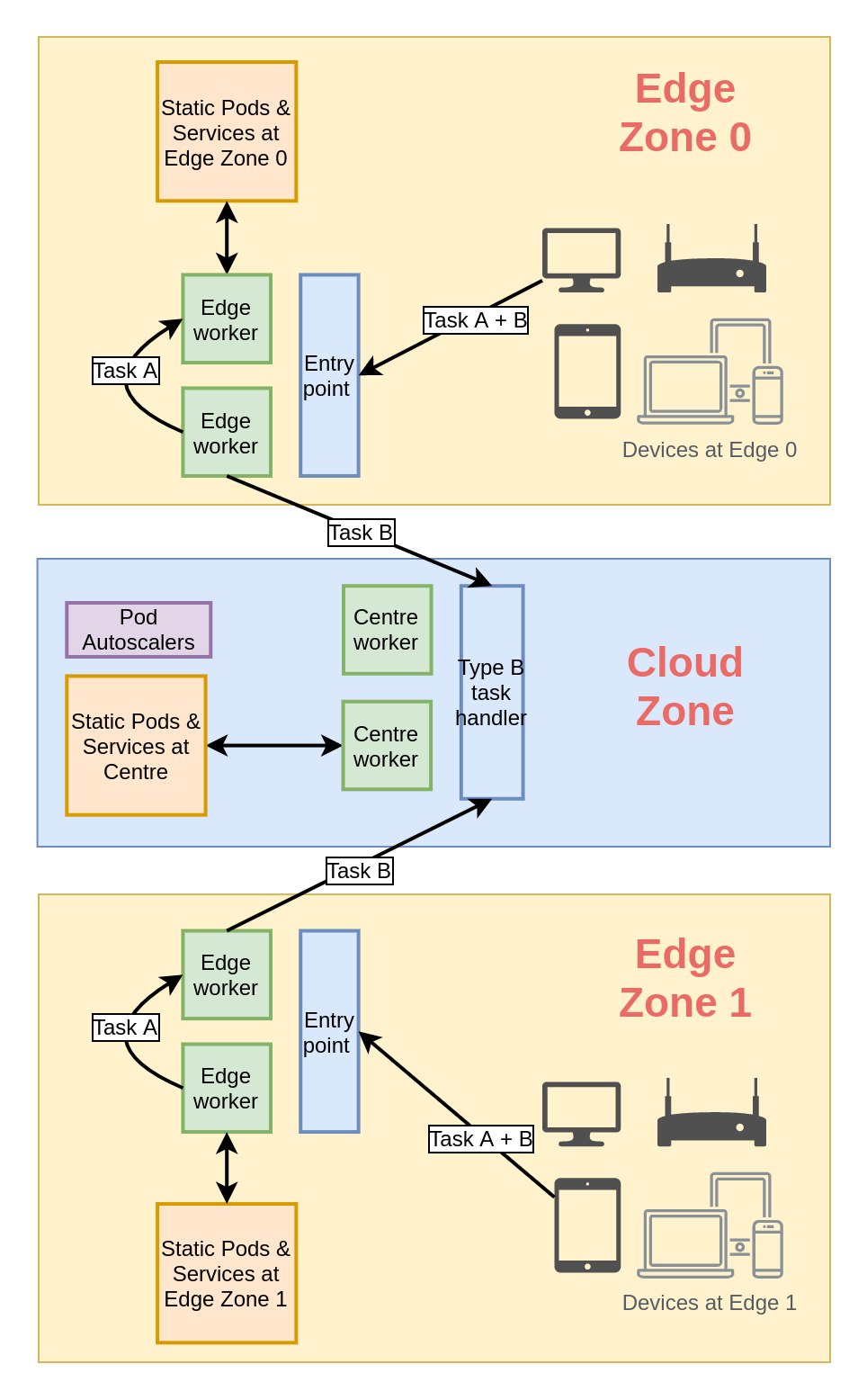}
    \caption{Architecture of example applications}
    \label{fig:app}
\end{figure}

\subsubsection{Topology}
Figure \ref{fig:app} shows the architecture of the example application hosted on the edge computing cluster. The example application is distributed in one cloud and two edge zones. At each zone, there deploy some supportive static pods and services, which are fixed and not designed to be scalable. In contrast, the worker pods in each zone are targets of autoscalers. All worker pods in every zone are providing a computing service and waiting for requests from clients or other pods. 

\subsubsection{Workflow \& Jobs}
All requests for the application are made from edge zones and reach entry points at their the edge closest to its location. Requests are classified as 2 different types, type A and type B, according to their handling cost. Type A requests are not costly, and are handled by edge workers. With larger computation cost, Type B requests will be forwarded to the cloud and completed in cloud. 

Here in the example application, Type A tasks are defined to sort a random array of a length of 3000 (referred as Sort tasks), while Type B tasks are to calculate eigen values of a matrix of dimension $(1000\times1000)$ (referred as Eigen tasks). The complexity of the Sort task is $n\log n$, which is $10^4$, while the complexity of the Eigen task is $n^3$, which is $10^9$. Sort tasks take less computation cost and they are handled by edge workers, while costs for Eigen tasks are larger and they are forwarded to and completed by cloud workers. 

 In both general and scientific computing areas, CPU-intensive applications are commonly used (e.g. weather forecasting, search algorithm, audio/video processing, etc.). The edge computing application used in this work is to emulate a typical CPU-intensive applications. Therefore, conclusions based on the example application are generalizable.

\subsection{Workload Generation}\label{sec:workload}
Two different workloads are considered in this work, namely \emph{Random Access} and National Aeronautics and Space Administration (\emph{NASA}) dataset. They are described in the text below.

\subsubsection{Random Access}
\emph{Random Access} is designed to generate workloads for applications randomly. The algorithm is shown as Algorithm \ref{alg:randaccess}. Both Sort and Eigen tasks are generated randomly with a probability of 0.9 and 0.1 respectively, to emulate that most requests which are less costly will be handled by edge workers while other tasks requiring large cost will be done in cloud. By accessing the application with 3 different patterns of workloads periodically, namely \var {light}, \var {medium}, \var{heavy}, the autoscaler is expected to cover most potential cases that may occur in real-world uses. 

\begin{algorithm}
\SetAlgoLined
\While{True}{
    \var {load\_type} $\leftarrow$ \var {Random([light, medium, heavy])}\;
    \var {request\_num} $\leftarrow$ \var {Random(Range(20, 200))}\;
    \For{\var i $\leftarrow$ \var 0 ; \var i $<$ \var {request\_num}; \var {i++}}
        {\var {task} $\leftarrow$ \var {Random([sort]*9 + [eigen])}\;
        \var {Request(task)}\;
        \var {sleep\_time\_range} $\leftarrow$
        \begin{equation*}
            \begin{cases}
                \var {Range(0.1, 0.3)} & \var {load\_type} $==$ \var {heavy}\\
                \var {Range(0.5, 1)} & \var {load\_type} $==$ \var {medium}\\
                \var {Range(2, 5)} & \var {load\_type} $==$ \var {light}\\
            \end{cases}
        \end{equation*}
        \var {Sleep(Random(sleep\_time\_range))}\\
    }
}
\caption{Algorithm of \emph{Random Access} for workload generation}
\label{alg:randaccess}
\end{algorithm}

\subsubsection{NASA dataset}
\begin{figure}[h]
    \centering
    \includegraphics[width=0.8\linewidth]{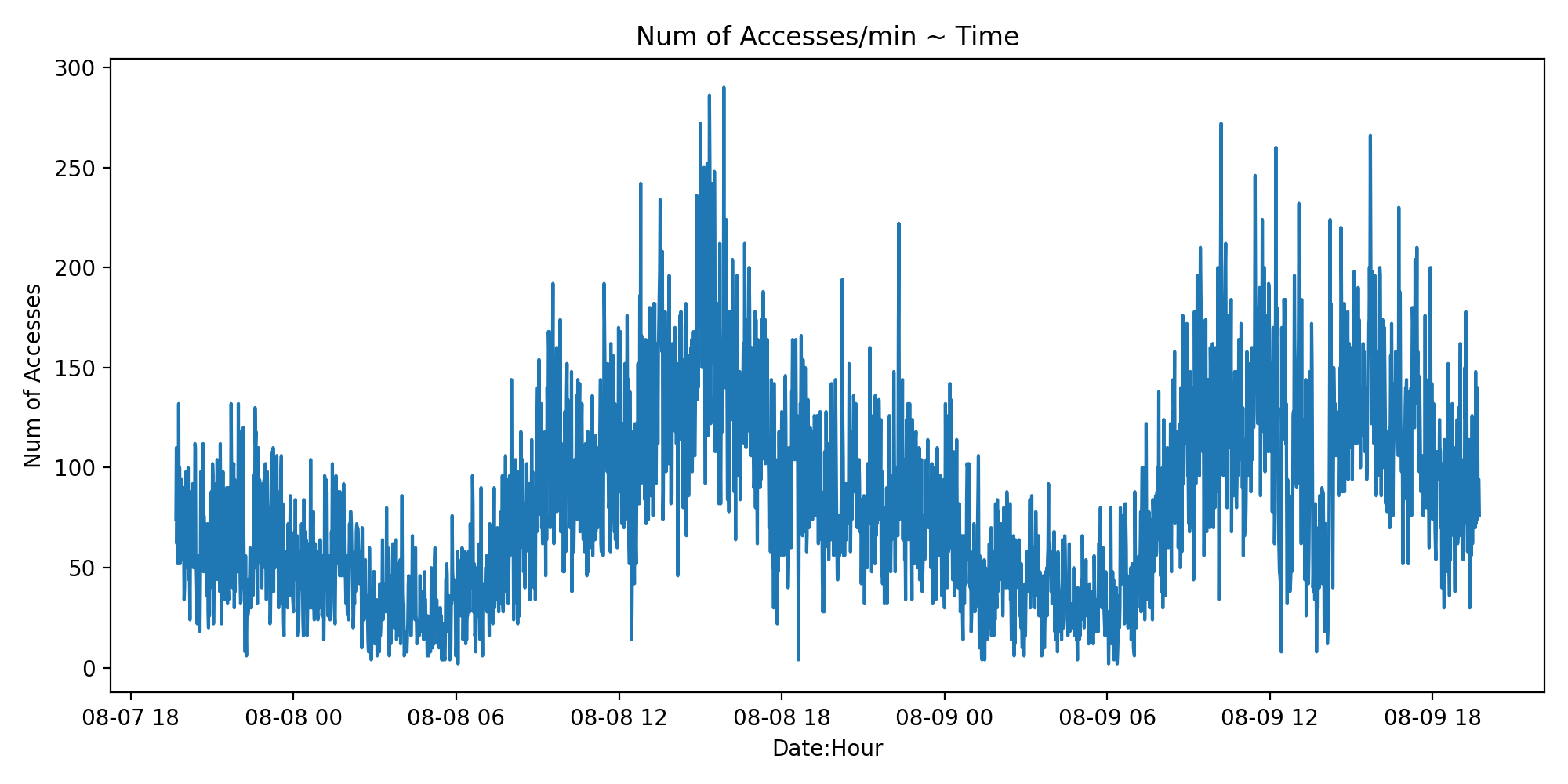}
    \caption{Subset of scaled \emph{NASA} requests to emulate real-world application accesses}
    \label{fig:nasa}
\end{figure}

\emph{NASA} dataset is used to emulate real-world access to the example application\cite{nasa1998}. All access records are collected by the NASA Kennedy Space Center WWW server in Florida. All requests are recorded as raw access logs in the form of binary files, in which each log contains its access timestamp. Raw log files are preprocessed by accumulating access requests for each minute, and aggregated number of logs are used to make requests to the application. 
A comprehensive set of experiments with all requests from the dataset, which contains records of 2 months would be very time consuming.
In this work, a subset of 2 days from the dataset are selected for experiments. Also, the number of requests are adjusted to a proper scale, so that the peak workload does not exceed resource limitations of the edge environment for experiments. Figure \ref{fig:nasa} shows the subset of scaled NASA dataset that is used in this work. Still, each request is randomly selected as a Sort or Eigen task by same probabilities as \emph{Random Access}. 

\subsection{Experiments for Optimization}
All experiments for the hyperparameter optimization of the proposed PPA are based on workloads generated by \emph{Random Access}. Three hyperparameters of the proposed PPA are to be optimized, including the time series model for workload predicting, Update Policy for the model and the key metric of the example application. The optimization problem is formulated as Equation \ref{eq:opt}. 
\begin{equation}\label{eq:opt}
\begin{aligned}
\min_{M, U,K} \quad & \gamma (M, U, K, A)\\
\text{and} \min_{M, U,K} \quad & W(M, U, K, A)\\
\textrm{s.t.} \quad & \sum_{p\in P_n} R_{p} \leq R_n \text{ for all } n \in N\\
\end{aligned}
\end{equation}
where $\gamma$ is the response time of the application, $W$ is the sum of wasted resources, $M$ is the predicting model, $U$ is the model update policy, $K$ is the key metric, $A$ is the target application, $R_p$ is the resource assigned to pod $p$, and $R_n$ is the resource capacity of node $n$. $P_n$ is the set of all pods hosted on node $n$ and $N$ is the set of all nodes. Here the target application $A$ is fixed, and $M, U, K$ are to be optimized to minimize $\gamma$ and $W$. 

\subsubsection{Optimization of Predicting Models}\label{sec:modelopt}
The predicting model used in the PPA is of vital importance for the performance of proactive pod autoscalers. However, for different applications and different workloads, optimal choices differ. In this experiments, an Auto Regressive Moving Average (ARMA) model and a Long Short-term Memory (LSTM) model are compared. The input metrics of both models are CPU, RAM, Network I/O utilization and HTTP requests rate and key metrics for both PPAs are set as CPU utilization. 

The ARMA model we use is ARMA(1, 1, 1), representing that orders of both its moving average and autoregressive part are set as 1, and time window of the model is 1. The formula for ARMA(1, 1, 1) is shown as Equation \ref{eq:arma11}, where $\mu$ is the mean of the series, $\theta_1$ is the parameter of the Moving Average part, $\varphi$ is the parameter of the Autoregressive part, and $\varepsilon$ is the error term. Hyperparameters of ARMA model is pre-selected based on collected data. 
\begin{equation}\label{eq:arma11}
        y_t = \mu + \varepsilon_t + \theta_1\varepsilon_{t-1} + \varphi_{1}y_{t-1}, 
\end{equation}

The LSTM model we use consists of a 50-neuron LSTM layer and a fully-connected layer activated by the ReLu function\cite{agarap2018deep}. The shape of the output layer is set as 5, to fit all future metrics. The loss function for the LSTM model is the Mean Squared Error and the optimizer is the Adam Optimizer\cite{kingma2014adam}. 

Dataset for pretraining both models is collected by running the example application for 10 hours with workloads generated by \emph{Random Access} on a single unconstrained node. The Dataset consisted of 1800 records, and among which, 1200 are used for pretraining the seed model, while others are for model validation. With different pretrained models injected, each PPA is used to autoscale the example application for 200 minutes, and predicted values and real values of CPU utilization are collected during the running time. 

\subsubsection{Optimization of the Update Policy}
As introduced in Section \ref{sec:updatePolicy}, 3 different approaches for the \emph{Updater} are proposed and compared. In this experiment, a LSTM model is pretrained and injected in the PPA as the seed model with the same settings in the experiment for model optimization, and CPU utilization is defined as the key metric.  To shorten the time required for the experiments, the time interval of model update loop is set to 1 hour, representing that the model is updated every hour. The example application runs for 200 minutes autoscaled by each PPA with different update policies, and predicted versus actual values of CPU utilization are collected.

\subsubsection{Optimization of the Key Metric}
For the example of the CPU-intensive application, either the request rate or the sum of CPU utilizations of all pods can be set as the key metric for PPAs, and in this experiment, both key metrics are compared. Like experiments for the optimization of other hyperparameters, example applications run for 200 minutes autoscaled by each PPA with workloads generated by \emph{Random Access}. In this experiment, due to the difference of key metrics, it is not possible to compare two PPAs quantatively with predicted metrics. Instead, response times of all requests and idle resources of the system autoscaled by two PPAs are used to quantify performance of two PPAs. 

\subsection{Experiments for Evaluation}\label{sec:exp}
With the optimal predicting model, update policy and key metric, the optimized PPA is yet to be evaluated in real-world scenarios. Experiments for the evaluation of the proposed PPA are conducted as follows. The application runs for 48 hours autoscaled by the PPA with optimal configurations. Workloads of the example application are generated by the scaled \emph{NASA} dataset. To evaluate performance of the application for users, response time of requests and idle resources are collected and compared. Another run of the application is conducted with exact same configurations autoscaled by a Horizontal Pod Autoscaler. This is the baseline for the proposed proactive pod autoscaler. 

\section{Results \& Discussion}
The results are presented and discussed in the following text. The hyperparameter optimization results are first presented, followed by evaluation in a deployed scenario.
\subsection{Optimization of the Predicting Model}
\begin{figure}[h]
    \centering
    \includegraphics[width=0.8\linewidth]{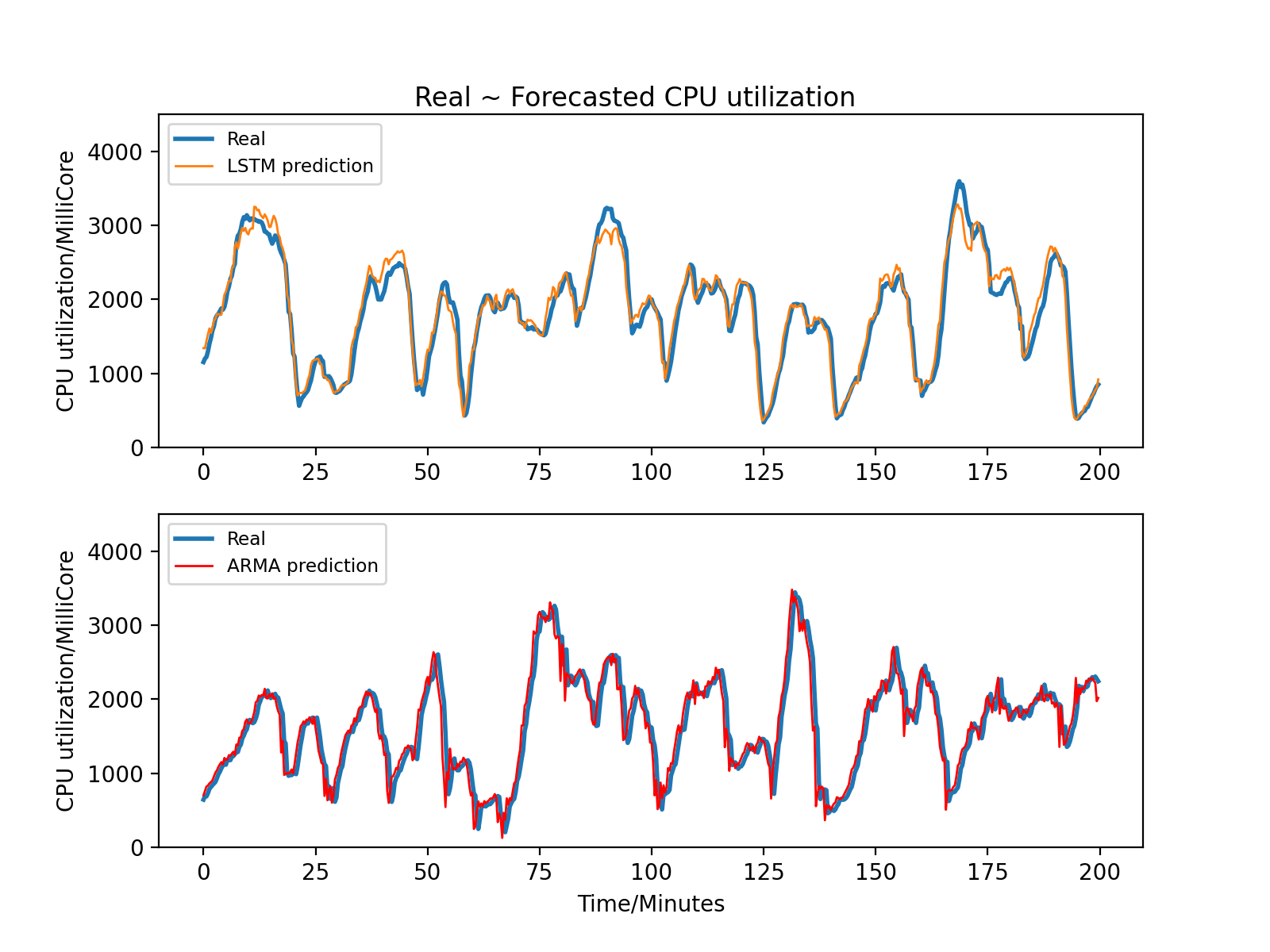}
    \caption{Comparison of predicting performance of the ARMA and LSTM model}
    \label{fig:res_real_model}
\end{figure}

The ARMA and LSTM model are compared by running example applications as real time use cases. Results of PPAs with two different models are plotted as Figure \ref{fig:res_real_model}. 
It can be observed that that both models are able to capture trends of CPU utilization, and the ARMA model is slightly better in terms of fitting. However, quantitatively, the Mean Squared Error (MSE) of predictions produced by the LSTM model is 53240.972 while that produced by the ARMA model is 96867.631, which is much larger. This indicates that ARMA model provides significant shifts when predicting CPU utilization and the LSTM model is able to produce relatively more accurate predictions. Therefore, it can be concluded that the LSTM outperforms the ARMA model for prediction of CPU utilization on the example application. 

\subsection{Optimization of the Update Policy}
\begin{figure}[h]
    \centering
    \includegraphics[width=0.8\linewidth]{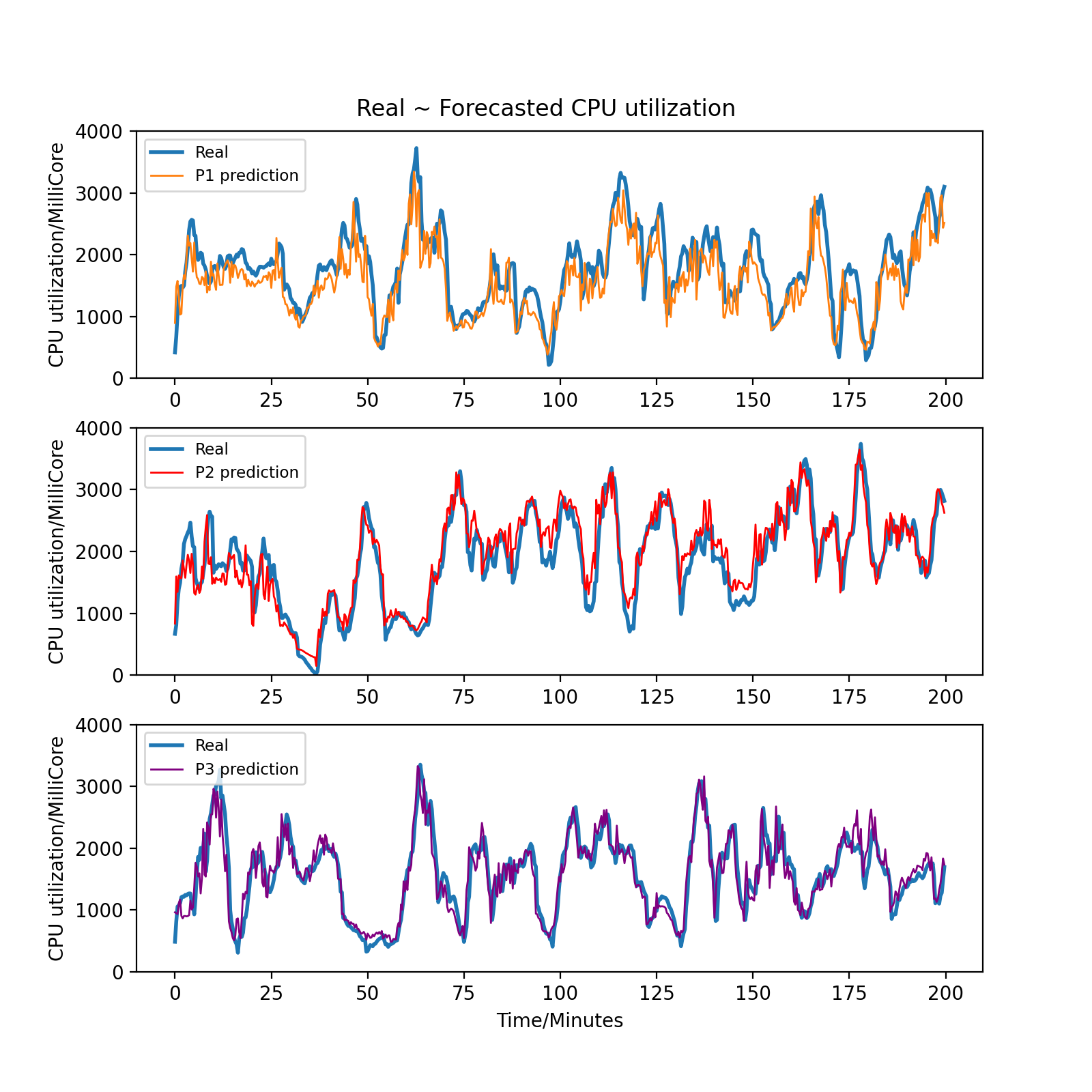}
    \caption{Comparison of predicting performance of different update policies on the real time data}
    \label{fig:res_real_policy}
\end{figure}

Three PPAs embedded with different policies are compared in running applications shown as Figure \ref{fig:res_real_policy}. It can be observed that Policy 1 performs worse than Policy 2 and 3, while Policy 3 provides better prediction accuracy than Policy 2. Quantitatively, the PPA with Policy 1, 2 and 3 produce utilization prediction with MSE of 64769.882, 42180.437 and 30994.449 respectively, which confirms that Policy 3 performs the best among 3 proposed model update policies. Therefore, we conclude that Policy 3, which retrains models for extra epochs with newly-collected metrics in each \emph{model update loop}, is able to provide the optimal performance for the example application over other policies. 

\subsection{Optimization of the Key Metric}
\begin{figure}[h]
    \centering
    \includegraphics[width=0.8\linewidth]{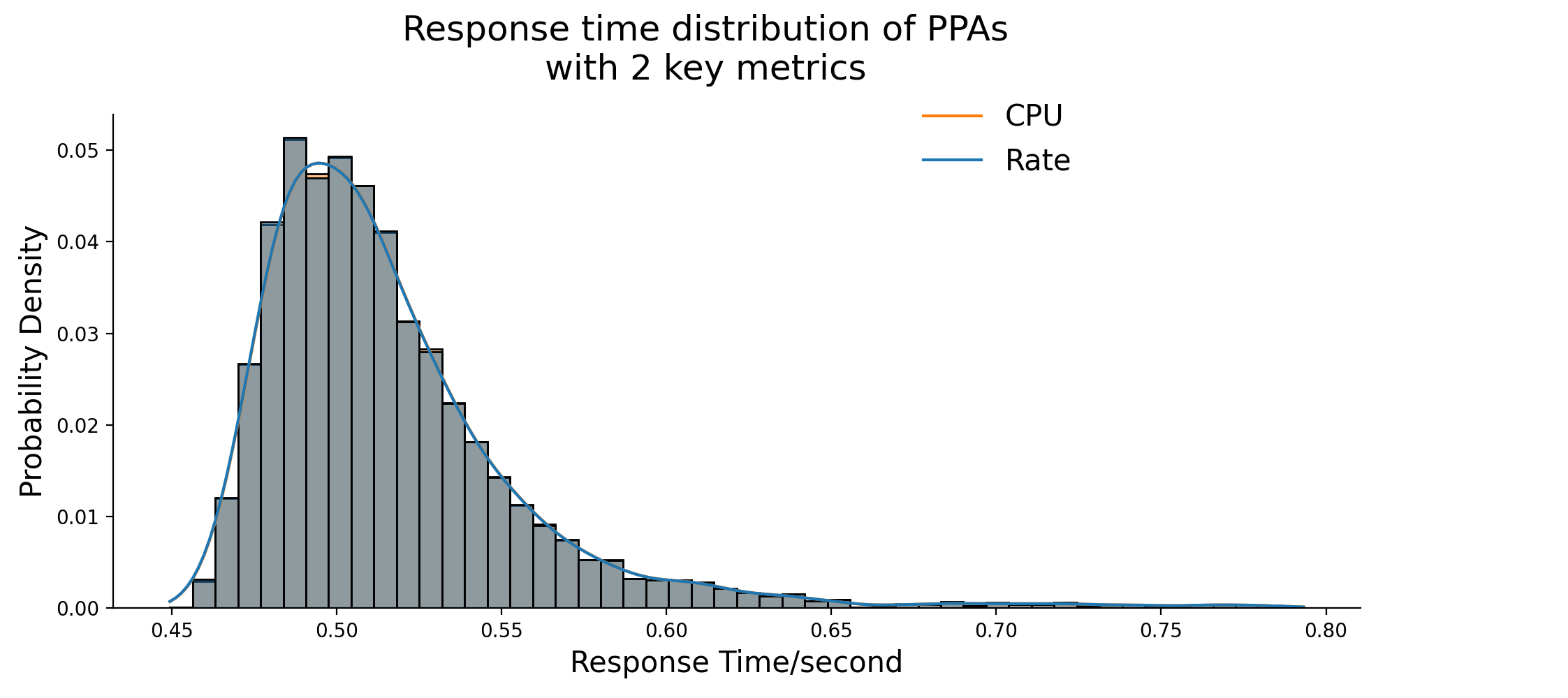}
    \caption{Comparison of response time of requests on applications autoscaled with differrent key metrics}
    \label{fig:res_metric_time}
\end{figure}

Two key metrics of the PPA, namely the rate of requests and CPU utilization, are compared on the example application. Response time of requests and idle resources are used to compare performance of PPAs quantitatively. Figure \ref{fig:res_metric_time} shows the response time distributions of requests on applications autoscaled with different key metrics. Huge overlaps of two distributions indicate that response time of two runs are extremely close. The average response time of the application autoscaled on CPU utilization is 0.5156s with a standard deviation (STD) of 0.0421, while the average response time of the application autoscaled on request rates is 0.5157s with a STD of 0.420. Two distributions are not significantly different and it can be concluded that two key metrics are equivalent when autoscaling the example application in terms of response time. 

\begin{figure}[h]
    \centering
    \includegraphics[width=0.8\linewidth]{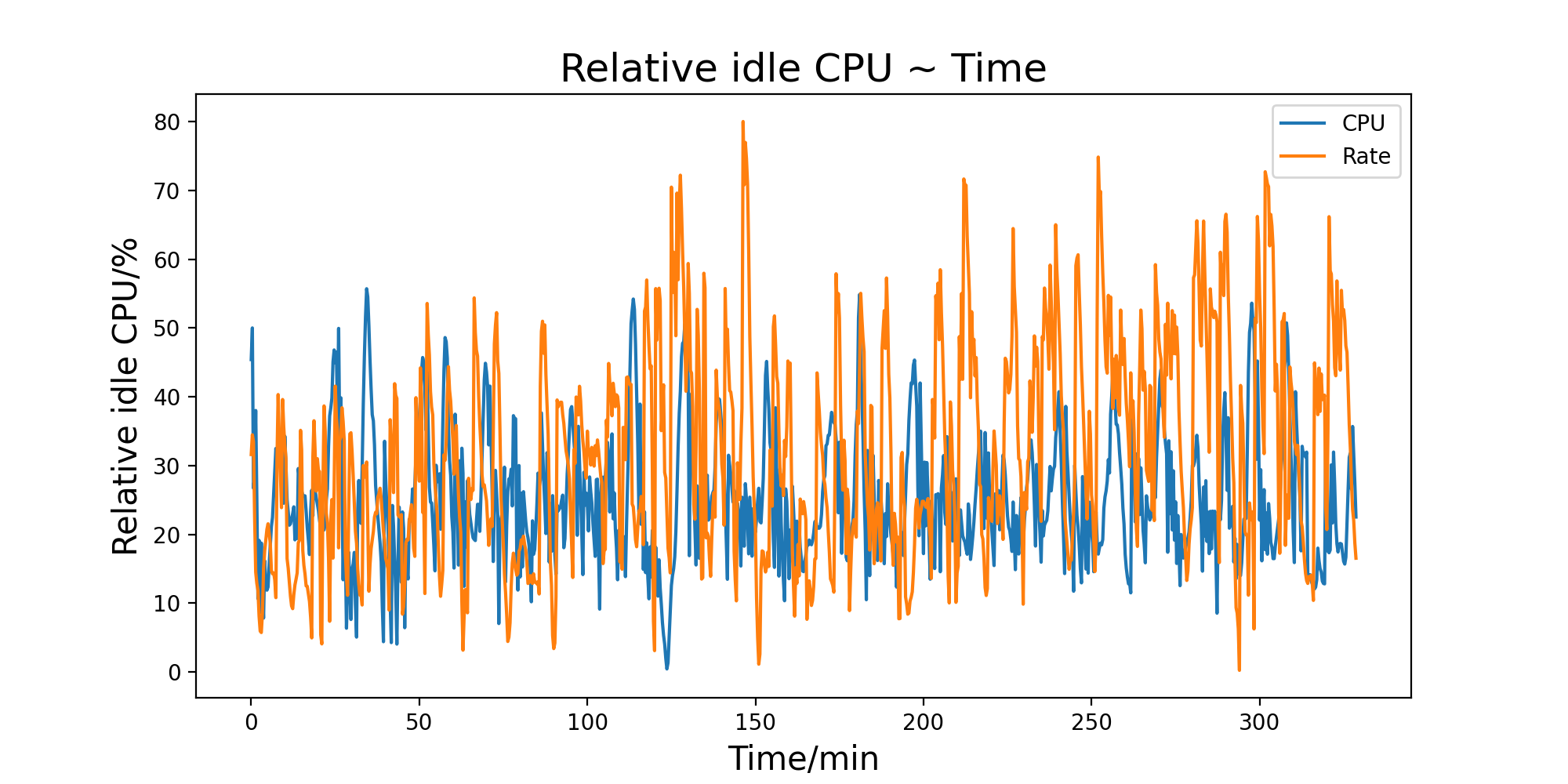}
    \caption{Comparison of relative idle resources on applications autoscaled with differrent key metrics}
    \label{fig:res_metric_resource}
\end{figure}

Beside response time, idle resources are compared to quantify amounts of resources wasted by two autoscalers. Here we define relative idle resources (\textit{RIR\_t}) of the system at time $t$ as Equation \ref{eq:rir}. RIRs by PPAs with different key metrics are plotted as Figure \ref{fig:res_metric_resource}. As we can observe, the application autoscaled by the PPA with the key metric of requests rate has larger RIR compared with the key metric of CPU utilization. Quantitatively, the average RIR of requests rate is 0.317 with a standard deviation of 0.161 while the average RIR of CPU utilization is 0.251 with a STD of 0.092. Thus, it can be concluded that the PPA with the key metric of CPU utilization is more efficient, wasting less resources. Moreoever, the lower standard deviation of RIR indicates that the system is more stable when autoscaled by CPU utilization as its key metric. 
\begin{equation}\label{eq:rir}
    \text{RIR}_t = \frac{\text{CPU\_idle}_t} {\text{CPU\_requested}_t }
\end{equation}

Though both PPAs with different key metrics are able to provide close response time for requests, the PPA with CPU utilization is more energy-efficient and the system is more stable. Therefore, for the example application, the PPA with CPU utilization is the optimal pod autoscaler. 

\subsection{Results of Experiments for Evaluation}
With the optimal model, key metric and update policy, the final evaluation of the proposed PPA is conducted as described in Section \ref{sec:exp} and performance of PPA and HPA are quantitatively compared with response time of requests and idle resources. 

\subsubsection{Response Time}
\begin{figure}[h]
    \centering
    \includegraphics[width=0.8\linewidth]{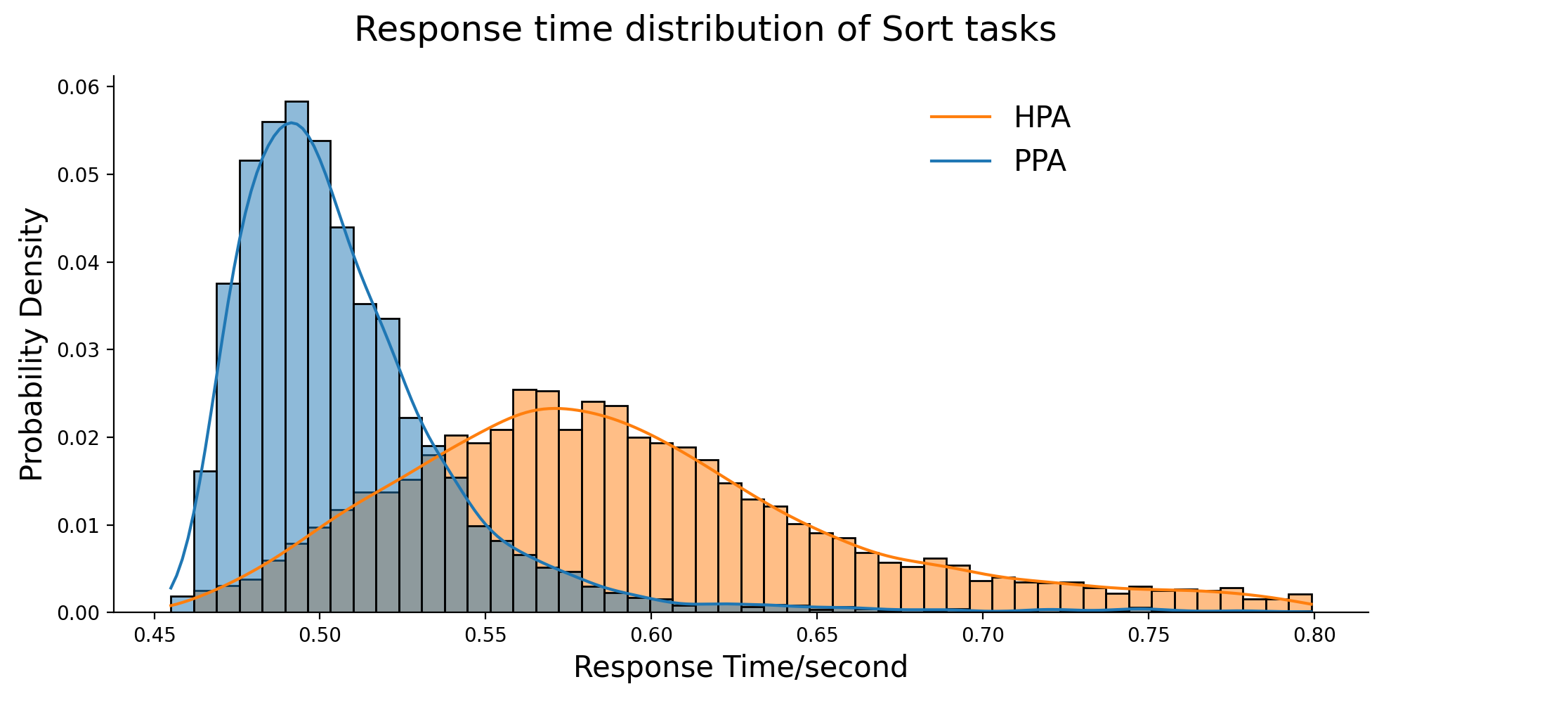}
    \caption{Comparison of response time distributions of Sort tasks autoscaled by HPA and PPA}
    \label{fig:eval_res_edge}
\end{figure}

Figure \ref{fig:eval_res_edge} shows distributions of response time for Sort tasks autoscaled by PPA and HPA. The average response time of the application autoscaled by HPA for edge tasks is 0.592 second with a standard deviation (STD) of 0.067, while it of the application autoscaled by PPA is 0.508 second with a STD of 0.038. The response time provided by PPA is significantly less than it of HPA with a p-value less than $10^{-3}$. Also, the distribution provided by PPA has a smaller standard deviation. It can be concluded that the application autoscaled by PPA provided less latency edge services and the edge system is more stable. 

\begin{figure}[h]
    \centering
    \includegraphics[width=0.8\linewidth]{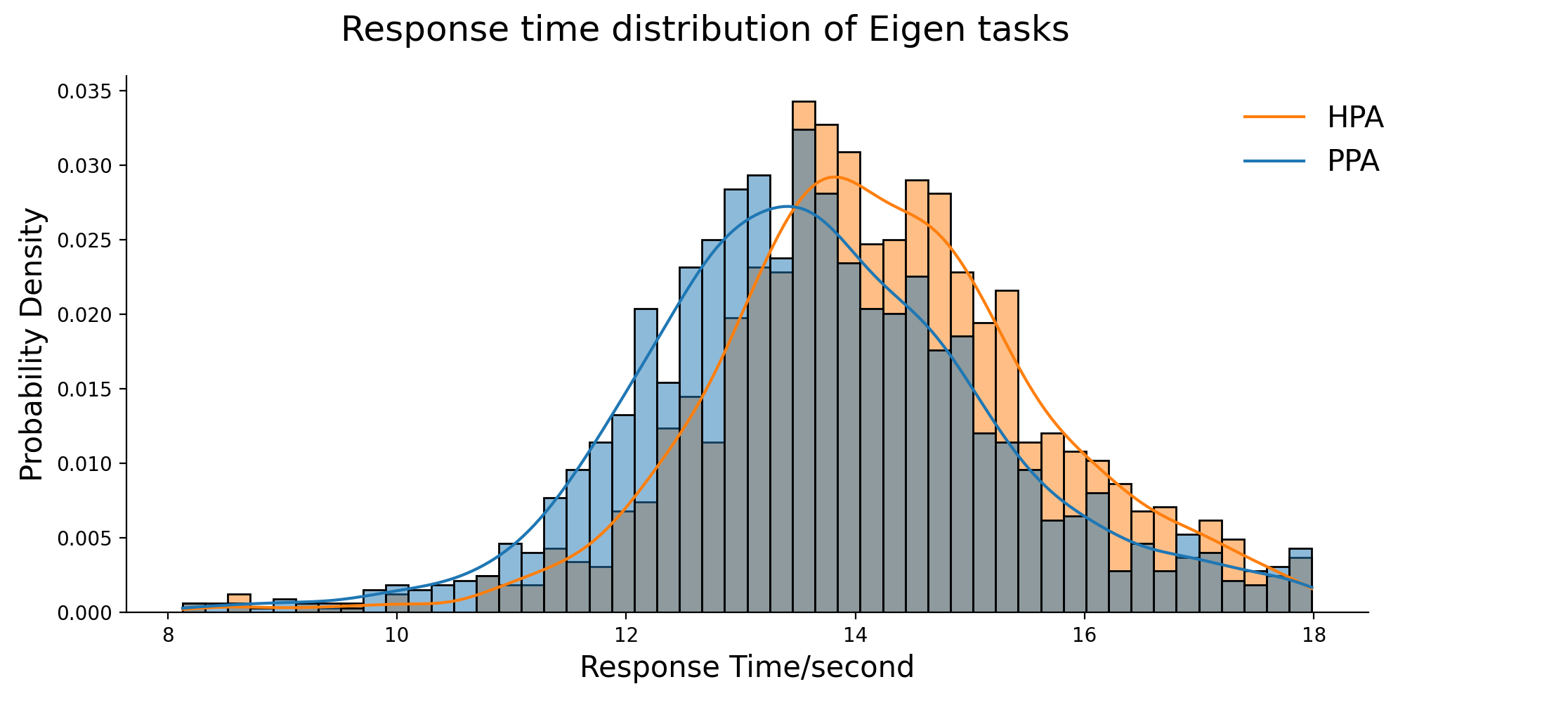}
    \caption{Comparison of response time distributions of Eigen tasks autoscaled by HPA and PPA}
    \label{fig:eval_res_cloud}
\end{figure}

Similar results are observed on Eigen tasks as well. Two distributions of response time for cloud tasks provided by HPA and PPA are shown as Figure \ref{fig:eval_res_cloud}. The average response time of the application autoscaled by HPA for edge tasks is 14.206 seconds with a standard deviation of 1.703, while it of the application autoscaled by PPA is 13.646 seconds with a STD of 1.576. Response time provided by HPA is signicantly larger than that of PPA with a p-value less than $10^{-3}$, and same holds for the standard deviation as for the Sort tasks. Thus, the cloud services autoscaled by PPA brings better performance in comparison to HPA. 

Therefore, it can be concluded that on edge computing applications, the proposed Proactive Pod Autoscaler outperforms the default Horizontal Pod Autoscaler for both cloud services and edge services, with lower latency and more access stability. 

\subsubsection{Idle Resources}
\begin{figure}[h]
    \centering
    \includegraphics[width=0.8\linewidth]{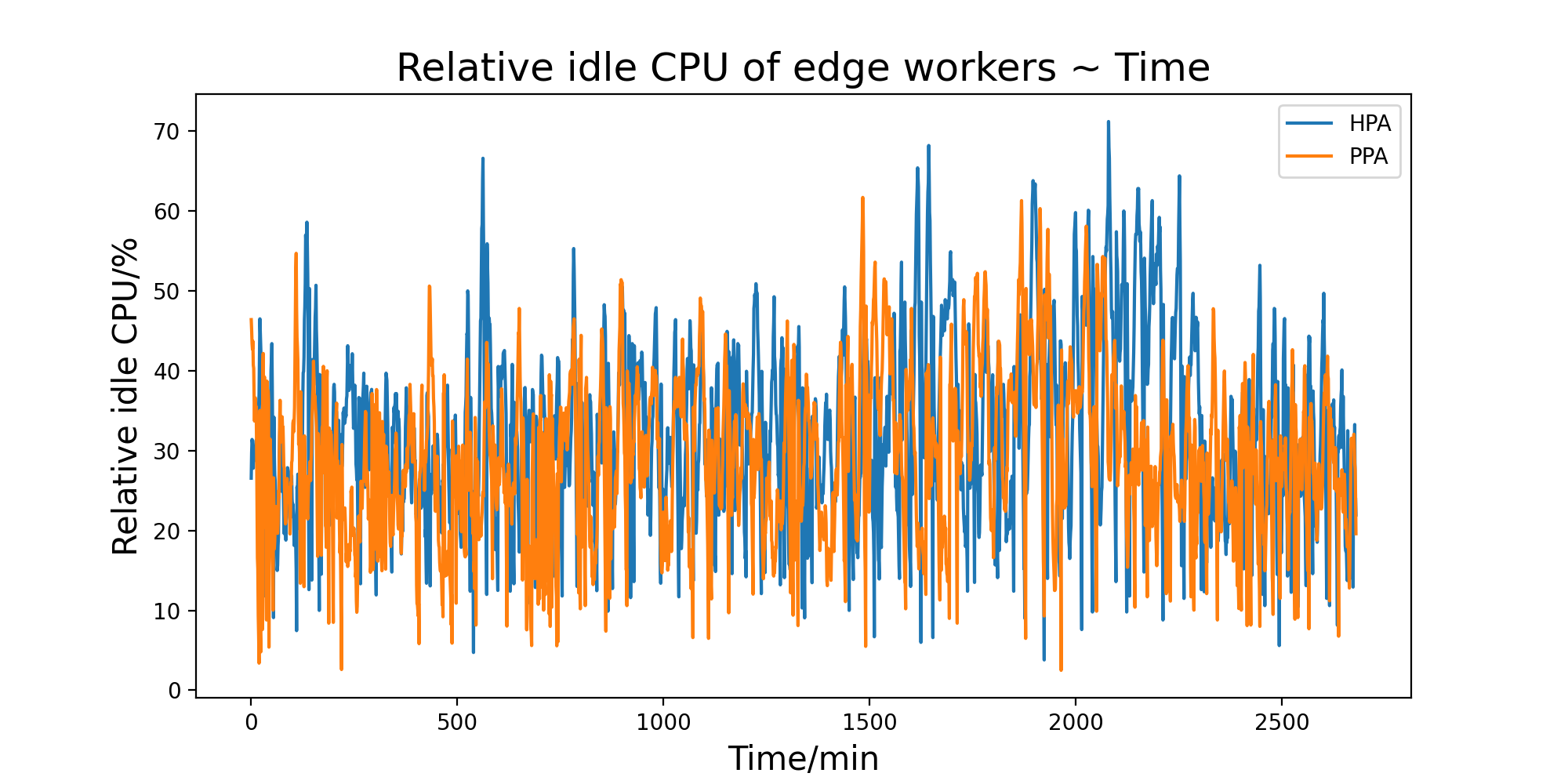}
    \caption{Comparison of relative idle CPU of edge workers autoscaled by HPA and PPA}
    \label{fig:eval_idle_edge}
\end{figure}

Figure \ref{fig:eval_idle_edge} shows relative idle CPU usage of edge worker nodes autoscaled by PPA and HPA. The average relative idle CPU of edge workers autoscaled by HPA is 0.3209 with a standard deviation of 0.1079, while that autoscaled by PPA is 0.2988 with a STD of 0.1026. Though results of two autoscalers are similar visually, the relative idle CPU provided by PPA is significantly less than that of PPA with a p-value of less than $10^{-3}$. Thus, it can be concluded that edge worker nodes of the application autoscaled by PPA utilize CPU resources more efficiently than the HPA autoscaled case. 

Relative idle CPU usage of cloud worker nodes autoscaled by HPA and PPA are shown as Figure \ref{fig:eval_res_cloud}. The average relative idle CPU of cloud workers autoscaled by HPA is 0.3373 with a standard deviation of 0.1572, while that of the application autoscaled by PPA is 0.3098 with a STD of 0.1453. Relative idle CPU provided by HPA is signicantly larger than that of PPA with a p-value of less than $10^{-3}$. Thus, on cloud nodes, the application autoscaled by PPA wastes less CPU resources than the application autoscaled by HPA as well. 

It can be concluded that in both terms of response time of requests and idle resources, the proposed Proactive Pod Autoscaler outperforms the default Horizontal Pod Autoscaler, bringing better performance for edge computing applications, gaining more access stability, and wasting less resources. 

\begin{figure}[h]
    \centering
    \includegraphics[width=0.8\linewidth]{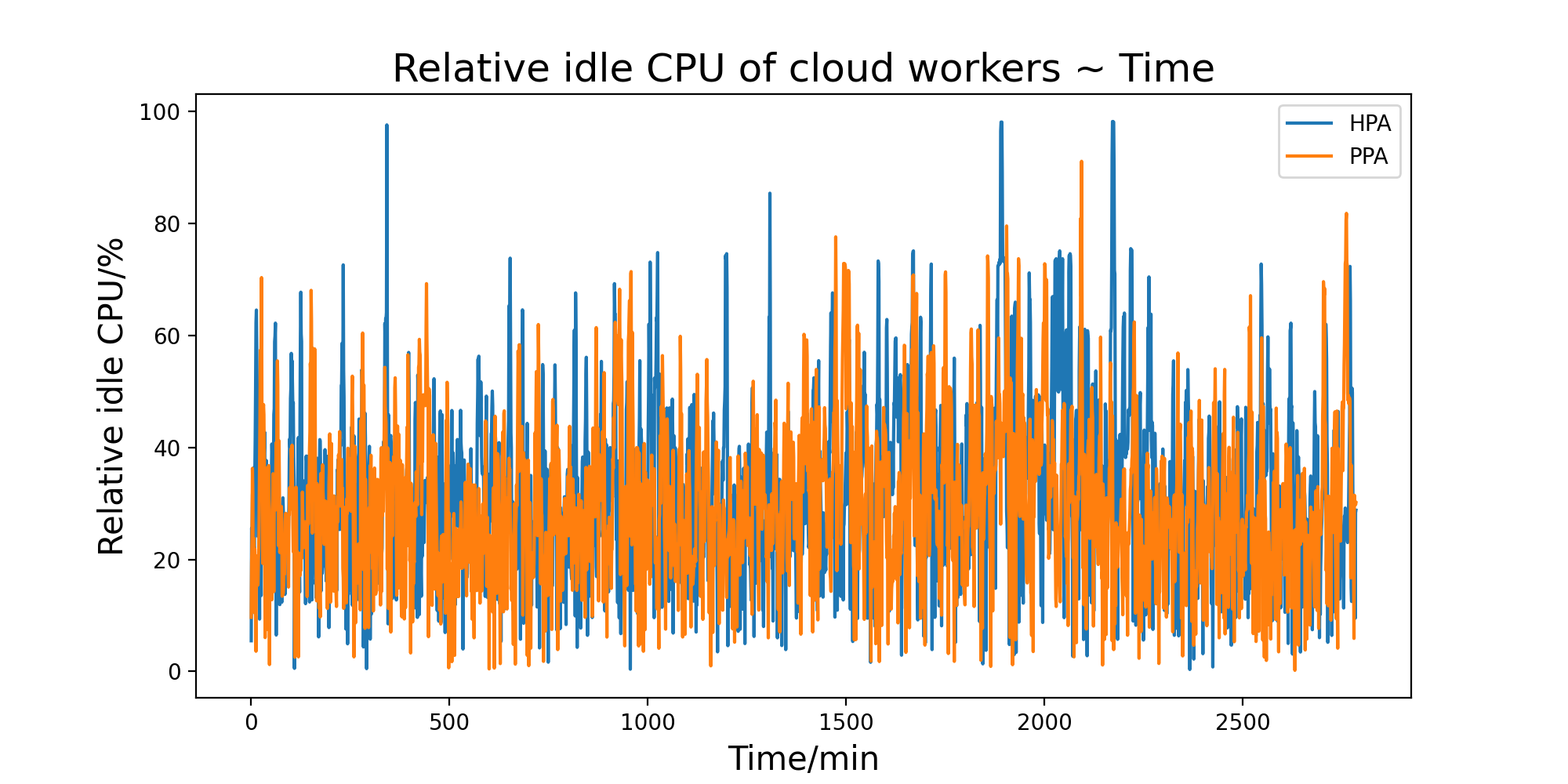}
    \caption{Comparison of relative idle CPU of cloud workers autoscaled by HPA and PPA}
    \label{fig:eval_idle_cloud}
\end{figure}

\section{Conclusion \& Future Work}
In this work, we propose a proactive pod autoscaling technique for edge computing applications based on Kubernetes. The proposed autoscaler is able to forecast arriving workloads of applications in advance using time series prediction methods, and to scale applications up and down as needed. Considering a CPU-intensive application as an example, the autoscaler is optimized and evaluated against the built-in autoscaler HPA in Kubernetes. It can be concluded that our proposed PPA is able to utilize resources more efficiently and to provide faster response times for requests. 

Besides, the proposed autoscaler is highly flexible and customizable, so that users can adjust metrics, predicting models and scaling policies to better fit needs of their own applications. In addition to multiple metrics being supported, users can also define their own application-specific metrics to better scale their applications. This enables PPAs to fit various types of applications, i.e. data-intensive applications, I/O-intensive applications in real-world uses. 

The proposed PPA is not perfect, since it requires developers to manually define optimal key metrics, models and policies to use. As future work, the PPA can be improved by automating the hyperparameter optimization involved. A possible approach may involve running the application with a set of possible metrics, with a designated module of the PPA modeling collected running data with different methods automatically. The best model can then be selected among candidate models using validation techniques. In this way, the auto-optimized PPA will be easier to deploy and use for developers. 

\section*{Acknowledgments}
The presented research is supported by the HASTE Project (Grant No. BD12-0008) and eSSENCE strategic collaboration. The authors also would like to acknowledge and thank Swedish National Infrastructure for Computing (SNIC) for providing cloud resources.

\bibliographystyle{unsrt}  
\bibliography{references}

\end{document}